\documentclass[prd,letterpaper,tightenlines,
showpacs,preprintnumbers,nofootinbib,floatfix,psfig]{revtex4}

\usepackage{amsmath,amssymb,amsfonts,graphicx,pifont,dcolumn}
\usepackage{epstopdf}
\usepackage{color}
\usepackage{cancel}
\usepackage{ulem}
\RequirePackage{slashed}

\begin{document}

\title{Scalar and Tensor Glueballs as Gravitons}

\author {Matteo Rinaldi}

\author{ Vicente Vento }

\address{Departamento de F\'{\i}sica Te\'orica-IFIC, Universidad de Valencia-
CSIC, 46100 Burjassot (Valencia), Spain.}

\date{\today}

\begin{abstract}
The bottom-up approach of the AdS/CFT correspondence leads to the study of
field
equations in an $AdS_5$ background and from their solutions to the 
determination of the hadronic 
mass spectrum. We extend the study to the equations of $AdS_5$ gravitons 
and determine from them the glueball spectrum. 
We propose an original presentation of the results which facilitates the
comparison of the various models with the spectrum obtained
 by  lattice QCD. This comparison  allows to draw some phenomenological 
conclusions. 

\end{abstract}


\maketitle

\section{Introduction}

Quantum Chromodynamics (QCD), the theory of the strong interactions, has eluded
an analytical solution since its formulation \cite{Fritzsch:1973pi}. One of the 
aspects of QCD which has attracted much attention is the glueball spectrum 
\cite{Fritzsch:1975wn,Mathieu:2008me}. In an attempt to understand this theory a 
procedure to extend the AdS/CFT correspondence breaking conformal invariance and 
supersymmetry was proposed 
\cite{Maldacena:1997re,Witten:1998qj,Witten:1998zw,Gubser:1998bc}.  In this so 
called top-down approach the glueball spectrum has been studied 
\cite{Csaki:1998qr,Constable:1999gb,Brower:2000rp,Vento:2017ice}. The AdS 
geometry of the dual theory is an AdS-black-hole geometry where the horizon 
plays the role of an infrared (IR) brane. 

One relevant feature found in ref.\cite{Brower:2000rp} is that the graviton of 
$AdS_7$, not the dilaton, corresponds to the lightest scalar glueball. This 
feature is in good agreement with the lattice QCD spectrum were the lightest 
scalar glueball has a much lower mass than its immediate excitation which is 
almost degenerate with the tensor glueball  
\cite{Mathieu:2008me,Bali:1993fb,Morningstar:1999rf,Vaccarino:1999ku,Lee:1999kv,
Bali:2000vr,Hart:2001fp,Lucini:2004my,Chen:2005mg,Gregory:2012hu,Lucini:2001ej}.
This observation motivates 
the present investigation.

A different strategy based of the AdS/CFT correspondence, the so-called
botton-up 
approach starts from QCD and 
attempts to construct a five-dimensional holographic dual. One implements 
duality in nearly conformal conditions defining QCD on the four dimensional 
boundary and introducing a bulk space which is a slice of $AdS_5$ whose size is 
related to $\Lambda_{QCD}$ 
\cite{Polchinski:2000uf,Brodsky:2003px,Erlich:2005qh,DaRold:2005mxj}. This is 
the so called hard-wall approximation. Later on, in order to reproduce the Regge 
trajectories, the  so called soft-wall approximation  was introduced 
\cite{Karch:2006pv,reggenew}. Within the bottom-up strategy and in both, hard-wall 
and the soft-wall approaches, glueballs 
arising from the correspondence 
of  fields in $AdS_5$ have  been studied 
\cite{BoschiFilho:2002vd,BoschiFilho:2005yh,Colangelo:2007pt,Forkel:2007ru,
Li:2013pta,Li:2013oda}. 

 In this scenario, the purpose of this investigation is to find the role 
of the $AdS_5$ graviton
in the bottom-up approach.  We study the spectrum of the scalar and tensor 
components of the $AdS_5$ graviton establishing a correspondence with the 
glueball spectrum of lattice QCD 
\cite{Morningstar:1999rf,Lucini:2004my,Chen:2005mg}  shown in Table 
\ref{masses}. 

\begin{table} [htb]
\begin{center}
\begin{tabular} {|c c c c c c c|}
\hline
& $0^{++}$&$2^{++}$&$0^{++}$&$2^{++}$&$0^{++}$&$0^{++}$\\
\hline
MP & $1730 \pm 94$ & $2400 \pm122$ & $2670 \pm 222 $&  & &  \\
\hline
YC & $1719 \pm 94$ & $2390 \pm124$ &  &  &  &  \\
\hline
LTW & $1475 \pm 72$ & $2150 \pm 104$ & $2755 \pm 124$& $2880 \pm 164 $& $3370 
\pm 180$& $3990 \pm 277$  \\
\hline
\end{tabular}
\caption{Glueball masses (MeV) from lattice calculations MP
\cite{Morningstar:1999rf}, YC \cite{Chen:2005mg} and LTW \cite{Lucini:2004my}. 
We have not 
included the lattice results from the unquenched calculation 
\cite{Gregory:2012hu} to be consistent, which are in agreement with the shown 
results within errors.}
\label{masses}
\end{center}
\end{table}

We will also show in our figures for completeness the $N \rightarrow \infty$ 
limit of the lightest scalar glueball.  The mass of the other glueballs do not
change much in this limit \cite{Lucini:2001ej,Lucini:2004my} as shown in table 
\ref{largeN}.

\begin{table}[htb]
\centering
\begin{tabular}{|c| c  c c c  c|}
\hline 
   m(SU(3))/m(SU($\infty$))& $0^{++}$ &  & $0^{++*}$ & & $ 2^{++}$\\ \hline 
  Continuum & $1.07\pm 0.04$ &   &$0.94  \pm 0.04$  &  & $1.00 \pm 0.03$
  \\ \hline
Smallest lattice  & $1.17 \pm 0.05 $&  & $1.00 \pm 0.04$ & & $0.99 \pm 0.04$
\\ \hline
\end{tabular}
\caption{Ratios of glueball masses for N=3 and very large N as shown in 
ref. \cite{Lucini:2004my,Lucini:2001ej}.}
\label{largeN}
\end{table}

In the next sections  we proceed to study the graviton in the hard-wall and
 soft-wall approaches to $AdS_5$ and compare the results
with previous calculations with scalar and tensor fields.  The graviton arises
from Einstein's equations, while the 
variational approach on the field Lagrangian gives rise to the equations of
motion for the fields. Thereafter we match the 
spectra obtained from these different models with the lattice QCD (LQCD)
glueball spectrum  
\cite{Morningstar:1999rf,Chen:2005mg,Lucini:2004my} and extract  
conclusions.

\section{Glueballs as hard-wall gravitons}

According to AdS/CFT correspondence massless scalar string states are dual to 
boundary scalar glueball operators 
\cite{Witten:1998zw,Gubser:1998bc,Polchinski:2000uf}. 
On the other hand scalar string excitations with mass $\mu$ couple to boundary
operators 
of dimension $\Delta= 2 + \sqrt{4+ (\mu R)^2} $. Glueball operators with spin
$J$ have 
dimension $\Delta= 4+ J$, thus a consistent coupling between string states with
mass $\mu$ 
and glueball operators with spin $J$ requires $(\mu R)^2= J(J+4)$. The glueball 
operators are massless and respecting conformal invariance. Once we introduce a 
size in the AdS space there is an infrared cut off in the boundary which is 
proportional to $1/\Lambda_{QCD}$, explicitly breaking conformal 
invariance. The presence of 
the slice implies an infinite tower of discrete modes for the bulk states. These 
bulk discrete modes are related to the masses of the non-conformal glueball 
operators.

In the bottom-up approach, supergravity fields in the $AdS_5$ slice times a 
compact $S_5$ space are considered an approximation for a string dual to QCD. 
The metric of this space can be written as

\begin{equation}
ds^2=\frac{R^2}{z^2} (dz^2 + \eta_{\mu \nu} dx^\mu dx^\nu) + R^2 d\Omega_5,
\label{metric}
\end{equation}
where $\eta_{\mu \nu}$ is the Minkowski metric and the size of the slice in the
holographic coordinate $0< z <  z_{max}$ is related to the scale of QCD,
$ z_{max} =\frac{1}{\Lambda_{QCD}}$.  The equation of motion for the scalar field with mass $\mu$ in $AdS_5$  
is obtained from the Lagrangian of a free scalar in the curved background  and leads to 
\cite{Witten:1998zw,Gubser:1998bc}

\begin{equation}
\partial_z^2 \Phi - \frac{3}{z} \partial_z \Phi + \eta_{\mu \nu} 
\partial^{\mu} \partial^{\nu} \Phi- \frac{(\mu R)^2}{z^2}
 \Phi= 0.
\label{scalar}
\end{equation}
Motivated by the work of ref.\cite{Brower:2000rp} in the top-down approach, we 
study the contribution of the scalar component of the massless graviton in the 
sliced $AdS_5$ geometry.  Writing the metric as 
$g_{a  b } = \bar{g}_{a b } +h_{a b}$ where 
$\bar{g}_{a  b}$ is the background metric Eq.(\ref{metric}),
which is a solution of Einstein's equations, we 
obtain for the perturbation $h$ linearizing Einstein's equations 

\begin{equation}
 -{1 \over 2} h_{a b; c}^{; c} - {1 \over 2} 
h^{c}_{c; a b} + {1 \over 2} h_{a c; b}^{; c}+ 
{1 \over 2} h_{b c; a}^{; c} + 4 h_{a b}=0,
\label{graviton}
\end{equation}
which are the field equations for the graviton. Choosing the gauge where the 
only non vanishing component is

\begin{equation}
 h_{t t} = (z^{-2}- z^2)\phi(z) e^{-m x_3}~,
\label{sgauge}
\end{equation}
where $m$ is the mass parameter. Substituting this ansatz into
Eq.(\ref{graviton})
we obtain Eq.(\ref{scalar}) for a plane wave solution 
$\Phi(x,z) \sim e^{-iP\cdot x} \phi(z)$ with $P^2=-M^2$ and $\mu R=0$,

\begin{equation}
\frac{d^2 \phi}{dz^2} - 3\frac{d \phi}{dz} +M^2 \phi = 0
\label{gscalar}
\end{equation}
Thus the scalar graviton 
equation is exactly the same equation 
as that for the massless scalar  fields  dual to the scalar glueballs 
\cite{BoschiFilho:2002vd,BoschiFilho:2005yh,Colangelo:2007pt}. 

Let us now  find the perturbation to Einstein's equations 
for the tensor component of the graviton. In this case we choose the gauge 

\begin{equation}
 h_{ij} = q_{ij} T(z) e^{-m x_3},
 \label{tgauge}
\end{equation}
being $i,j=1,2$ and $q_{ij}$ a generic constant traceless-symmetric matrix. 
The result
of this calculation for $T(z)$  is also Eq.(\ref{gscalar}). Thus the scalar 
graviton and tensor 
graviton components lead to the same equation. The two components are 
degenerate.

The spectrum of the 
graviton coincides with that of the scalar fields and the tensor graviton with
the tensor fields if we add a mass term $(\mu R)^2 = 12$ if we use 
the same boundary conditions. Let us recall these solutions.

The  plane wave solutions

\begin{equation}
\Phi (x_{\mu},z) = C_k e^{-iP x} z^2 J_2(u_k z),
\label{scalarsolution}
\end{equation}
 with the following  boundary conditions,  
\begin{eqnarray}
\mbox{Dirichlet  } & J_2( \chi_k) & = 0 , \nonumber \\
\mbox{Neumann} & J_1( \xi_k) &=0 ,
\label{modeequations}
\end{eqnarray}
determine the scalar spectrum. Here $J_2$ and $J_1$ are  Bessel functions, 
for the Dirichlet modes 
$ u_k=\chi_k \Lambda_{QCD}$, while for the Neumann modes $u_k=\xi_k 
\Lambda_{QCD}$ and $k$ labels the energy modes.
The corresponding solutions for the mass of the glueballs in dimensionless units
 of  $ \Lambda_{QCD}$ are given by the zeros of  the corresponding Bessel 
functions. The energy modes of the scalar glueball are shown in Table 
\ref{modesold}. 

\begin{table} [htb]
\begin{center}
\begin{tabular} {|c c c c c c c|}
\hline k & 1 & 2 & 3 &4 &5& ...\\  
\hline
D scalar $\,$ & 5.136 & 8.417 & 11.620 & 14.796 &17.960 &... \\ 
\hline
N scalar & 3.832 & 7.016 & 10.173 & 13.324& 16.471 &... \\ 
\hline
\end{tabular}
\caption{Energy modes for the scalar glueball in the hard-wall model
with Dirichlet (D) and Neumann
(N) boundary conditions.}
\label{modesold}
\end{center}
\end{table}
\vskip -0.2cm

For the tensor modes according to duality we simply have to add the mass term
$(\mu R)^2 =J(J+4)$. Considering 
again plane wave solutions the spectrum is given by ,

\begin{center}
\vskip 0.2cm
\begin{tabular}{ l c c c }
Dirichlet & $J_n ( \chi_{ n,k}) = 0$  , \\
Neumann & $ (2-n) J_{n} ( \xi_{n,k}) + \xi_{n,k} J_{n-1} ( \xi_{n,k}) =0$ ,
\end{tabular}
\end{center}
\vskip -0.8cm
\begin{equation}
\label{modeequationsT}
\end{equation}
where $J_n$ are Bessel functions, $ n = 2+J $ and $k$  labels  the modes.

We show  the tensor modes in Table {\ref{modesTnew}  
\cite{reggenew,BoschiFilho:2005yh}. 

\begin{table} [htb]
\vskip 0.2cm
\begin{center}
\begin{tabular} {|l c c c c c c|}
\hline
\hspace{1.4cm} k & 1 & 2 & 3 &4 & 5 & ...\\
\hline
D tensor  & 7.588 & 11.065 & 14.373 & 17.616 &20.827&... \\
\hline
N tensor & 5.981 & 9.537 & 12.854 & 16.096&19.304& ... \\
\hline
\end{tabular}
\caption{Energy modes for the tensor field in the hard-wall model 
with Dirichlet (D) and 
Neumann (N) boundary conditions.}
\label{modesTnew}
\end{center}
\end{table}

We proceed to compare this $AdS_5$ spectrum to  the quenched 
LQCD glueball spectrum \cite{Morningstar:1999rf,Lucini:2004my,Chen:2005mg} in 
Fig.\ref{hardwall}.  To this aim, we fix the
scale of the $AdS_5$ calculation  by performing a best fit to the 
lowest glueball state. Once the $AdS_5$ spectrum is plotted with this initial 
scale we 
seed the data into the plot attributing the mode numbers to the different 
glueball states. Finally we modify slightly the scale of our original fit to get 
a best fit to the data. Note that we plot two masses for the lightest glueballs,
which correspond to the ones obtained by 
LQCD and its large $N$ limit as shown in Table \ref{largeN}  
\cite{Lucini:2001ej,Lucini:2004my}.

\begin{figure}[htb]
\begin{center}
\includegraphics[scale= 0.7]{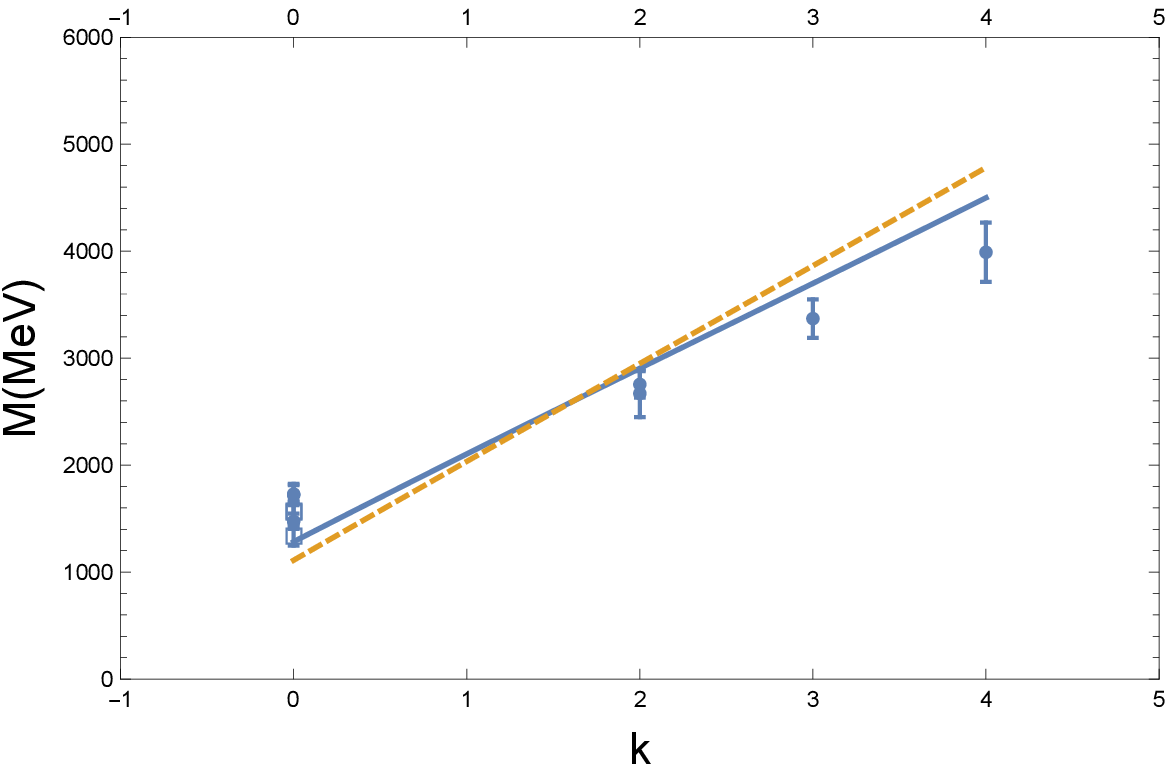} 
\hskip 0.5cm 
\includegraphics[scale= 0.7]{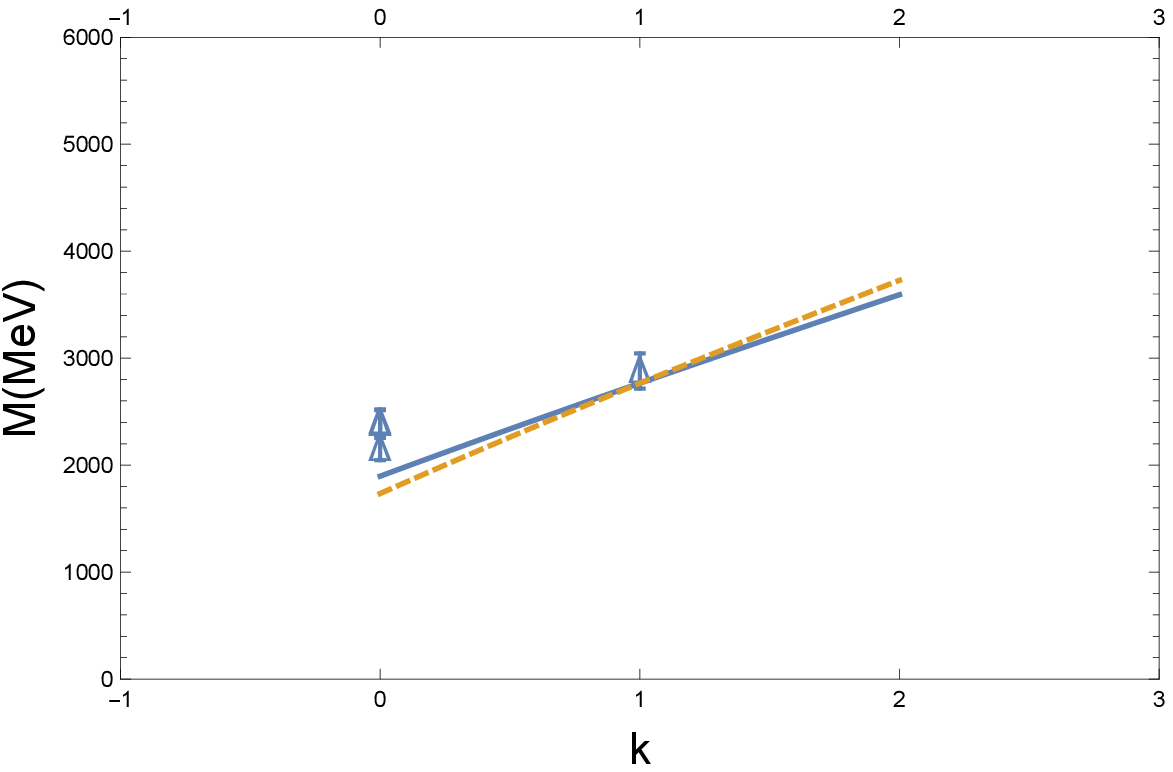} 
\end{center}
\caption{Gluebal spectrum obtained in the hard-wall model. The figure on the
left shows the fit to the scalar (J=0) glueball spectrum.  The figure on the
right shows fit to the tensor ($J=2$) glueball spectrum. 
The solid lines correspond to 
Dirichlet boundary conditions and the dashed lines to Neumann boundary 
conditions.
The full circles represent the scalar LQCD masses, the squares the large N 
limit scalar LQCD masses  and the 
triangles the tensor LQCD masses.}
\label{hardwall}
\end{figure}
The figure to the left shows the Dirichlet and Neumann fit to the scalar 
spectrum. We have skipped the 
$k=1$ mode because its value is $2713$ MeV is too high for a reasonable fit.
The figure on the right shows 
the fit to the tensor glueballs which has been obtained incorporating the mass
term. Note that since the scale is the same for scalar and tensor we have to 
get
a best fit to both data sets.

While the relation between the scalar graviton component and the scalar 
glueball is 
analogous to that of the scalar field: same equations and no $AdS_5$ mass, the
tensor graviton is not unless 
we introduce artificially a mass term. However the LQCD spectrum is telling us
that the second tensor glueball 
and the second scalar are almost degenerate. Moreover by looking at the modes
in Tables \ref{modesold} and \ref{modesTnew}
one realizes that the second scalar mode and the first tensor modes are almost
degenerate, very much so in the Dirichlet case.
Guided by this almost degeneracy of the scalar and tensor LQCD masses and of 
the degenerate scalar and tensor modes
we plot both scalar and tensor modes  following the massless equation, since 
the graviton 
equations for  scalar and tensor are degenerate.  To get a reasonable fit to 
the data, 
we skip for the tensor the $k=0$ mode, i.e., the lowest tensor glueball at
$2313$ 
MeV is ascribed to $k=1$ and since the next scalar glueball, the $2713$ MeV, is
almost 
degenerate with the $2880$ MeV tensor glueball we assign to it the $k=2$ mode 
number.   The result is shown in  Fig. \ref{hardwallST}. The fit is quite good 
with only one curve for scalar and tensor glueballs. 
The mode skipping in the case of the tensor is like a mass gap as we shall see 
in the soft-wall models.

\begin{figure}[htb]
\begin{center}
\includegraphics[scale= 0.7]{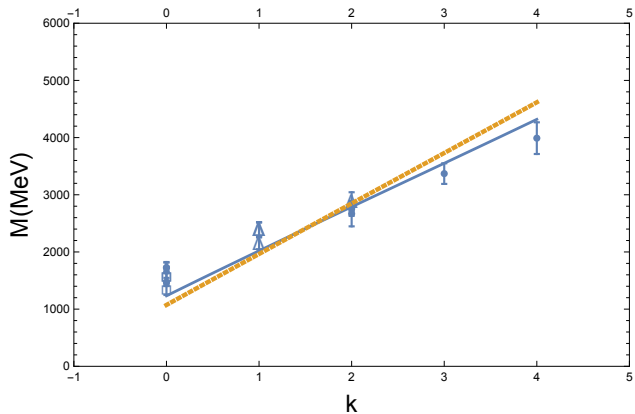} 
\end{center}
\caption{Gluebal spectrum obtained in the hard-wall model. We plot the modes
of the massless equation which 
represents the degeneracy of the scalar and tensor gravitons
skipping the $k=0$ mode for the tensor and the $k=1$ mode for the scalar. 
The solid lines correspond to 
Dirichlet boundary conditions and the dashed lines to Neumann boundary 
conditions.
The full circles represent the scalar LQCD masses, the squares the large N 
limit scalar LQCD masses  and the 
triangles the tensor LQCD masses.}
\label{hardwallST}
\end{figure}

\section{Glueballs as soft-wall gravitons}

Another scheme to determine the spectrum of QCD from $AdS_5$ has been a 
mechanism for a gravitational background which cuts-off smoothly in the 
holographic coordinate. The mechanism introduced some time ago for that 
purpose, capable of 
reproducing the Regge behavior of mesons, consists in 
incorporating a dilaton field $\delta$ and a metric $g_{MN}$ with characteristic 
properties \cite{Karch:2006pv}. In this formalism the glueballs are described  
by 5d fields propagating in this background with the action given by 

\begin{equation}
\mathcal{I} = \int{ d^5x \sqrt{- g} e^{-\delta} \mathcal{L}} ,
\label{lagrangian}
\end{equation}
where  $\mathcal{L}$ is the  lagrangian density describing the dynamics of the scalar fields,  
$\delta$ the dilaton whose mission is to soften the $z$ cut-off,
\begin{equation}
\delta (z) = \beta^2 z^2.
\end{equation}
and $g = \det{g_{a b}}$, where $g_{a b}$ is the $AdS_5$ metric defined in
Eq. (\ref{metric}) \cite{Colangelo:2007pt}. 
Since our aim  is to find the glueballs associated with the $AdS$ graviton  
without changing the results for conventional hadrons, we generalize the 
metric to  

\begin{equation}
g_{a b} (z)= e^{-\alpha^2 (z^2/R^2)} \frac{R^2}{z^2}  (-1,1,1,1,1).
\label{metric2}
\end{equation}
Eq. (\ref{metric2}) is a 
modification of the metric suitable for 
enriching the 
dynamics  
 of the $AdS$ graviton. These type of metrics have been used to explore heavy 
 quark 
 physics \cite{Andreev:2006ct,White:2007tu,Bruni:2018dqm}.
 In order to implement the condition of Regge 
trajectories used in previous calculation \cite{Karch:2006pv,Colangelo:2007pt} 
we must impose

\begin{equation}
\frac{  3 \alpha^2}{2} + \beta^2  =1.
\label{Colangelo}
\end{equation}
Note that the change of metric affects the lagrangian term in 
the action leading to an additional multiplicative factor $e^{\alpha^2z^2}$ in 
the integral which is the reason for the factor $3/2$ in Eq. (\ref{Colangelo}).
Such a choice 
for the metric produces a Lagrangian  for a scalar field that is identical to 
that of 
Eq. (\ref{lagrangian}) and will lead to the same equations
of motion
 for the fields.  We add no new degrees of freedom. However, the graviton 
 equations of motion change.

Given these restrictions, the spectra for the glueballs, arising from the 
scalar 
and tensor fields, using $R=1$ in Eq. (\ref{metric2}), are those of ref. 
\cite{Colangelo:2007pt},  i.e., 
$ M^2_J = 4k +4 +2\sqrt{ 4 + J(J+4)}, \; k=0,1, \ldots$, for even $J$. We 
rewrite these equations for scalar and tensor  in a 
single expression, 

\begin{equation}
M^2 =4k + 8, 
\label{Colangelomodes}
\end{equation}
where $k= 0,1,...$ for the scalar modes and  $k=1,2,..$ for the tensor modes.
One should notice  that the addition of the mass 
term to the tensor is equivalent to skipping 
the $k=0$ mode. In this model the phenomenological procedure for the spectra
discussed for the hard-wall model is perfectly realized.
These solutions correspond also to the graviton with the old metric Eq. 
(\ref{metric}), i.e. $ \alpha=0$

Let us find the spectrum for the scalar component of the graviton by solving
the Einstein's equations corresponding to the new metric Eq. (\ref{metric2}).
Using  $R=1$ and the same gauge Eq.(\ref{sgauge}) the mode equation becomes

\begin{equation}
\frac{d^2 \phi}{d z^2}  + \left(\alpha^2 z - \frac{3}{z}\right) \frac{d \phi}{d 
z} +
\left(\frac{8}{z^2} + 6 \alpha^2 + M^2 + 4 \alpha^2 z^2\right) \phi - 
\frac{8}{z^2} e^{-\alpha^2 z^2} \phi =0.
\end{equation}
Note that for $\alpha^2=0$ this equation reduces to Eq.(\ref{gscalar}). From 
this one can obtain a Schr\"odinger type equation,

\begin{equation}
- \Psi''(z) +V_{GS}(z) \Psi(z) = M^2 \Psi(z),
\label{GS}
\end{equation}
where 
\begin{equation}
V_{GS}(z)= \frac{8 e^{\alpha^2 z^2}}{z^2} - \frac{17}{4 z^2} -7 \alpha^2  - 
\frac{15 \alpha^4 z^2 }{4}.
\label{VGS}
\end{equation}

 In terms of the new 
variable $\tau= \alpha z/\sqrt{2}$, Eq.(\ref{GS})
becomes

\begin{equation}
-  \frac{d^2 \Psi}{d \tau^2} (\tau) + V_{GSS} (\tau) \Psi (\tau) = 
\Lambda^2 \Psi (\tau),
\label{tauSch}
\end{equation}
where 

\begin{equation}
\Lambda^2= \frac{2 M^2}{\alpha^2 } 
\label{modes}
\end{equation} 
and the potential is given by

\begin{equation}
V_{GSS}(\tau)= \frac{8 e^{2 \tau^2}}{\tau^2} - \frac{17}{4 \tau^2} -14  -15
\tau^2.
\label{pot0tau}
\end{equation}
In order to study the boundary conditions at the origin we Taylor expand the 
exponential to second order, obtaining a  
potential whose behavior at small $\tau$ is 

\begin{equation}
V_{low}(\tau) \sim \frac{15}{4 \tau^2}~, 
\end{equation}
which leads to a low $\tau$ behavior for the field function

\begin{equation}
\Psi(\tau) \sim \tau^{5/2}.
\end{equation}
Note that $\alpha$ is just a scale in the mass equation and 
that the mode solutions, $\Lambda_k$, are independent of 
its value. For small values of $\tau$ this equation leads to that of 
ref.\cite{Colangelo:2007pt}  for the scalar field up to an irrelevant constant
$\Lambda_0^2$.

For $\alpha^2>0$, the potential is not binding and the corresponding  
solutions for the eigenfunctions are damped oscillations. 
The well  behaved modes appear for $\alpha^2 <0$. In this case we use
the   variable $t= i\tau, t>0$ and $a= i\alpha, a^2>0$. 
Eqs.(\ref{tauSch}), (\ref{modes}) and (\ref{pot0tau}) now become

\begin{equation}
-  \frac{d^2 \Psi}{d t^2} (t) + V_{GSS} (t) \Psi (t) = \Lambda^2 \Psi (t),
\label{tauSch1}
\end{equation}
where,  

\begin{equation}
\Lambda^2= \frac{2 M^2}{a^2} 
\label{modes1}
\end{equation}
 and

\begin{equation}
V_{GSS}(t)= \frac{8 e^{2 t^2}}{t^2} - \frac{17}{4 t^2} +14  -15 t^2.
\label{pot0}
\end{equation}
It is important to note the change of sign in the exponential but also in the
constant term which lead to quantized modes solutions.
We show in Fig. \ref{solutions} the eigenfunctions for the first three modes. 
The corresponding eigenvalues,  $\Lambda_k$},  are given in Table 
\ref{sLambdamodes}.  The change of $\tau$ by $t$ corresponds 
to the change in the asymptotic behaviour to go from a non bound  to a bound 
solution.

For tensor component of the graviton  choosing the same gauge  
Eq.(\ref{tgauge}) one obtains  the same 
equation as for the scalar component Eqs. (\ref{GS}) and (\ref{VGS}).

\begin{figure}[htb]
\begin{center}
\includegraphics[scale= 1.0]{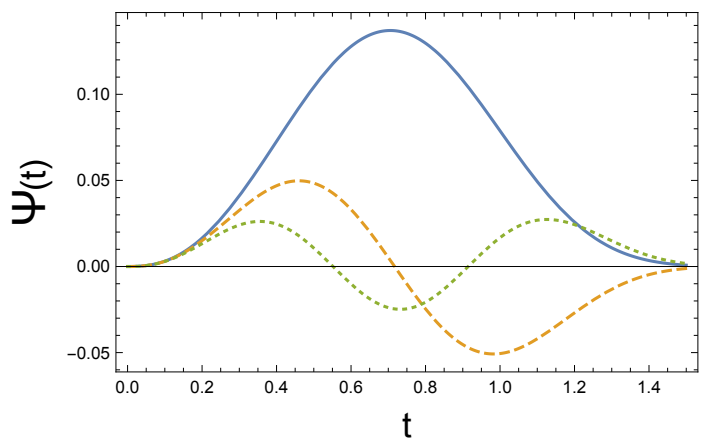} 
\end{center}
\caption{Typical eigenfunctions  of the  graviton 
equation for $\alpha^2 <0$. The solid curve shows the lowest (k=0) 
mode, the dashed curve that of the first mode (k=1) and the dotted curve that of 
the the second mode (k=2).} 
\label{solutions}
\end{figure}

\begin{table} [htb]
\begin{center}
\vskip 0.2cm
\begin{tabular}{| c c c c c c |}
\hline
k & 0 & 1 & 2& 4 & $\ldots$\\
\hline
scalar graviton & 7.341 &9.065 & 10.818 & 12.568&$\ldots$\\
\hline
\end{tabular}
\caption{The scalar modes $\Lambda_k$ of the graviton equation in the 
soft-wall model.}
\label{sLambdamodes} 
\end{center}
\end{table} 

In Table \ref{massmodes} we show the corresponding mass modes $M_k$ for scalar
and tensor fields \cite{Colangelo:2007pt}
and scalar and tensor graviton components calculated above.   
\begin{table} [htb]
\begin{center}
\vskip 0.2cm
\begin{tabular}{| c c c c c c |}
\hline
k & 0 & 1 & 2& 4 & $\ldots$\\
\hline
scalar field & 2.82  & 3.46 & 4.00 & 4.47 &$\ldots$\\
\hline
tensor field & 3.46  & 4.00 & 4.47 & 4.89 &$\ldots$\\
\hline
scalar \& tensor graviton & 5.19 &  6.41 &  7.65& 8.89&$\ldots$\\
\hline
\end{tabular}
\caption{The mass modes $M_k$ for the scalar,  tensor and the scalar and tensor 
graviton components in the 
soft-wall model.}
\label{massmodes} 
\end{center}
\end{table} 

Let us compare the results of the two soft-wall models studied with the
LQCD results. 
In order to perform such comparison, we   fix the scale of the 
$AdS_5$ 
calculation  by performing a best fit to the lattice data of 
the scalar glueball. Our fit requires a value
of $|\alpha| \sim 0.32$ which is close to $0.375$ obtained in ref. \cite{Brodsky:2007hb} by studying the pion form factor in a soft-wall model, and somewhat smaller that that of ref. \cite{Andreev:2006ct} , $0.47$, obtained by analyzing heavy quark dynamics also in a soft-wall model.

\begin{figure}[htb]
\begin{center}
\includegraphics[scale= 0.65]{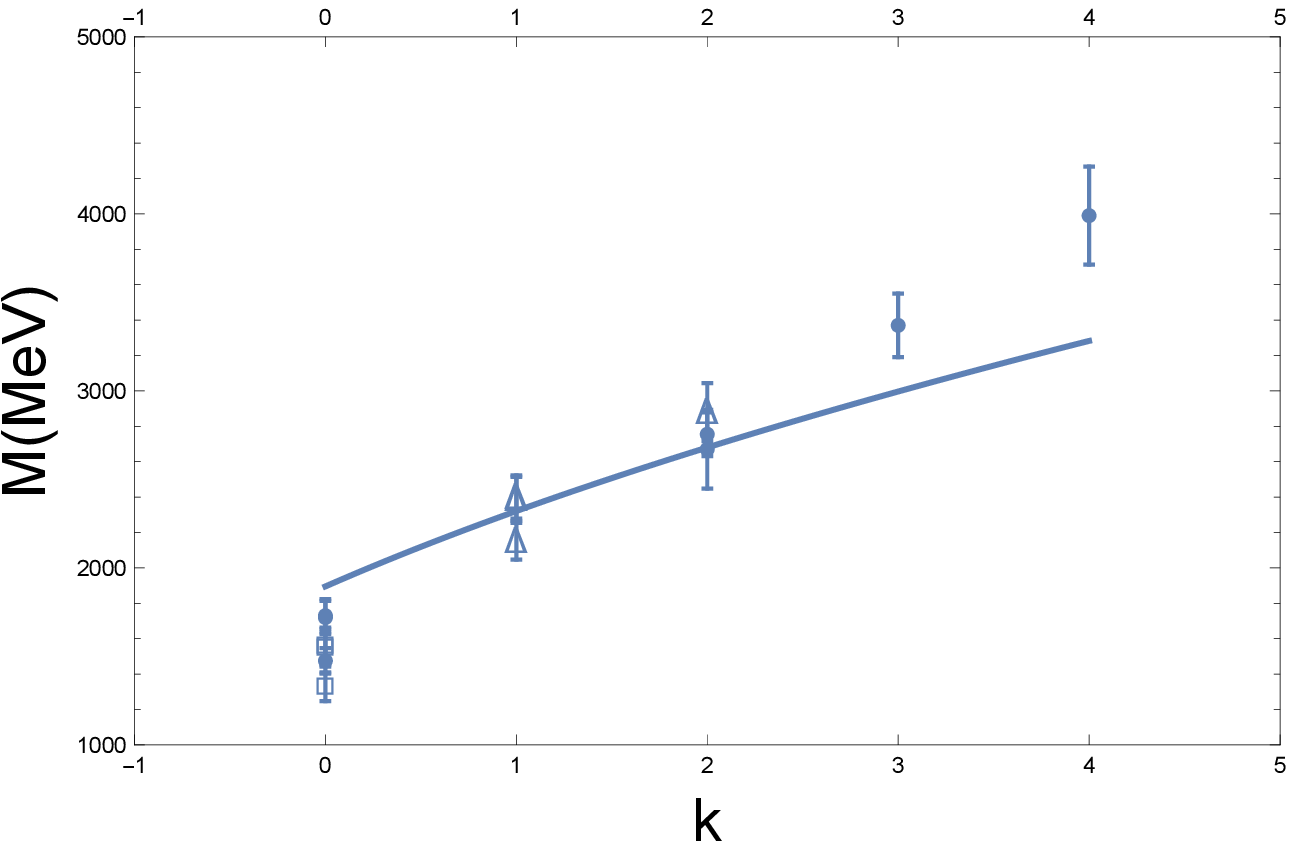} 
\hskip 0.5cm 
\includegraphics[scale= 0.65]{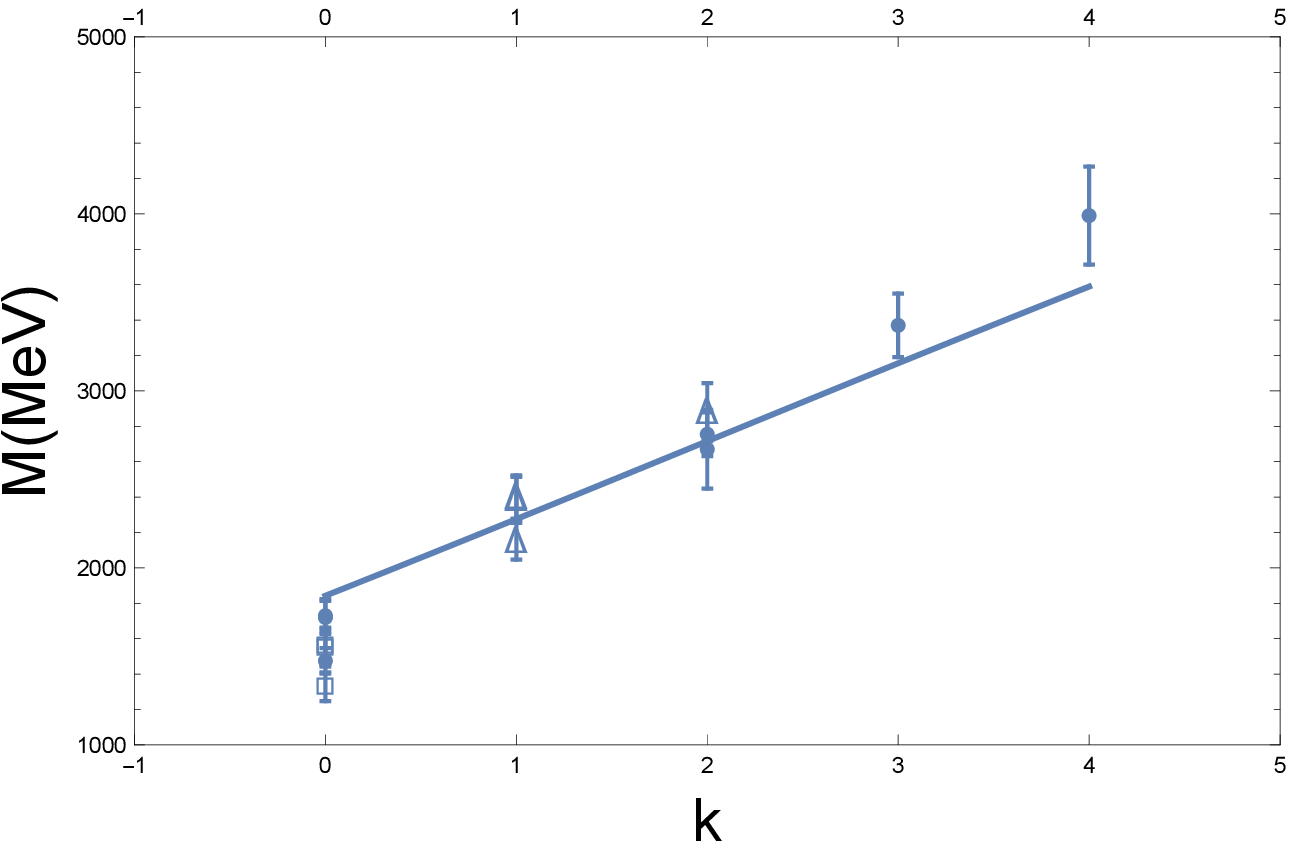} 
\end{center}
\caption{Glueball spectrum obtained by the soft-wall approach.  The figure on 
the left shows the 
scalar and tensor field results. The figure on the 
right shows the graviton fit to the glueballs. The symbols as in previous 
figures} 
\label{softwall}
\end{figure}

In Fig.\ref{softwall} we plot the graviton and field results for the glueball 
spectrum, by fitting the scale to the scalar spectrum, and compare them
to  the  
LQCD spectrum. We plot again two masses for the lightest glueballs, the 
one 
obtained by  LQCD and its large $N$ limit. For the fields (left)  we see 
 by looking at  Eq. (\ref{Colangelomodes}) that the scalar equation 
gives 
the tensor spectrum simply by shifting $k$ in one unit as result of adding a 
mass to the tensor. From $k=1$ on, the $AdS_5$
spectrum of 
scalar and tensor are degenerate. However, by looking at the lattice 
spectrum we notice that the degeneracy appears for the $k=2$ 
tensor, thus we ascribe $k=2$ to the second scalar. An important result of the 
analysis is as before the missing of a $k=1$ scalar.  The overall fit is 
reasonable 
but of lower quality than the hard-wall fit with Dirichlet boundary conditions.
For the gravitons (right) we proceed as 
before, a strategy that is now justified by the 
field equations. We skip the 
$k=0$ tensor mode and ascribe the first tensor mode to $k=1$. This could be 
understood as a way of implementing the tensor mass in the graviton approach. 
Since the scalar and the tensor components of the graviton are also  degenerate 
we ascribe $k=2$ to the second scalar. The fit is of better quality than the 
conventional 
soft model approach and the almost linear behavior describes better the data. 
It is clear that 
a further modification of the metric in line with  
refs. \cite{White:2007tu,Bruni:2018dqm} is
needed to get the adequate slope.  
 
\begin{figure}[htb]
\begin{center}
\includegraphics[scale= 0.75]{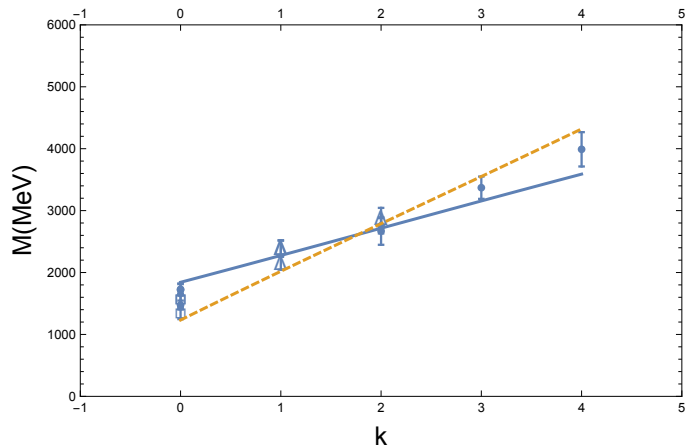} 
\end{center}
\caption{Gluebal spectrum obtained by the soft-wall (solid) and hard-wall 
(dashed) approaches.  The symbols as in previous figures} 
\label{softhard}
\end{figure}

Finally we compare following the same scheme described above the soft graviton 
with the hard Dirichlet graviton in Fig \ref{softhard}. 
Both give reasonable fits to the data although their slopes are quite different. 
The slope in the hard-wall model is too large, while that of the soft-wall model 
is too small. Thus both require more sophisticated 
metrics to describe better the data. 

\section{Conclusions}
We have discussed the spectrum of the scalar and tensor glueballs under the 
assumption that in an $AdS_5$ approach scalar and tensor components of the 
graviton might play a significant role corresponding to the lowest lying 
glueballs. We have studied the problem in hard and soft-wall models.

In the hard-wall model \cite{BoschiFilho:2002vd, BoschiFilho:2005yh} the scalar 
component of the graviton reproduces exactly the same equations as the field 
approach but the tensor component is degenerate. By fitting an energy 
scale the results of the model reproduce reasonably well the lattice results, 
specially so 
with Dirichlet boundary conditions.
The almost linear behavior of the fit is in good agreement with the data.

In the soft-wall approach we study a modification of the dilaton model 
\cite{Colangelo:2007pt} with a different metric leading to equations for the 
graviton which are not the same as those for the field equations. We find 
new solutions which 
depend on a metric parameter $\alpha^2 <0$. The metric grows as
$e^{-\alpha^2z^2}$ which implies that 
it grows at short distances becoming pure $AdS_5$ at infinity leading to a 
potential which is able to bind. We have solved numerically the equations for 
the scalar component. The tensor component turns out to be degenerate unless a 
mass term is added.
Both the field and graviton fits to the QCD lattice spectrum reasonable if the
first tensor mode is
ascribed to $k=1$ and the scalar mode at $k=1$ is skipped. The exact solutions
of the field equations give 
an explanation for the missing $k=0$ tensor mode 
if we interpret the effect as a tensor mass. In the case of the graviton this 
missing mode can be interpreted as faking a tensor mass. 
In both cases  we see that the doubling required by $AdS_5$ of the $k=1$ 
glueball is 
missing. We consider this a prediction of both soft-wall models. If this state 
 does not appear the equivalent $AdS_5$ dynamics at low mass
has to be more complicated. 
The graviton solution seems to better reproduce 
the shape and rise of the lattice glueball spectrum. 

The main conclusion of this paper is that we do not need to introduce additional fields into any $AdS_5$ model, the gravitons, 
with the addition of mass terms to satisfy the duality boundary conditions, are able to describe the elementary scalar and tensor glueballs. Fields might be useful to 
describe more complicated glueball structures. 

\section*{Acknowledgments}
We thank Marco Traini and Sergio Scopetta for discussions.
This work was supported in part by Mineco and UE Feder under contract FPA2016-77177-C2-1-P, 
GVA- PROMETEOII/2014/066 and SEV-2014-0398.


\begin{thebibliography}{99}

\bibitem{Fritzsch:1973pi}
  H.~Fritzsch, M.~Gell-Mann and H.~Leutwyler,
  Phys.\ Lett.\  {\bf 47B} (1973) 365.

\bibitem{Fritzsch:1975wn}
  H.~Fritzsch and P.~Minkowski,
  Phys.\ Lett.\  {\bf 56B} (1975) 69.

\bibitem{Mathieu:2008me}
  V.~Mathieu, N.~Kochelev and V.~Vento,
  Int.\ J.\ Mod.\ Phys.\ E {\bf 18} (2009) 1
  [arXiv:0810.4453 [hep-ph]].

\bibitem{Maldacena:1997re}
  J.~M.~Maldacena,
  Int.\ J.\ Theor.\ Phys.\  {\bf 38} (1999) 1113
   [Adv.\ Theor.\ Math.\ Phys.\  {\bf 2} (1998) 231]
  [hep-th/9711200].
  
\bibitem{Witten:1998qj}
  E.~Witten,
  Adv.\ Theor.\ Math.\ Phys.\  {\bf 2} (1998) 253
  [hep-th/9802150].

\bibitem{Witten:1998zw}
  E.~Witten,
  Adv.\ Theor.\ Math.\ Phys.\  {\bf 2} (1998) 505
  [hep-th/9803131].

\bibitem{Gubser:1998bc}
  S.~S.~Gubser, I.~R.~Klebanov and A.~M.~Polyakov,
  Phys.\ Lett.\ B {\bf 428} (1998) 105
  doi:10.1016/S0370-2693(98)00377-3
  [hep-th/9802109].

\bibitem{Csaki:1998qr}
  C.~Csaki, H.~Ooguri, Y.~Oz and J.~Terning,
  JHEP {\bf 9901} (1999) 017
  [hep-th/9806021].
  
\bibitem{Constable:1999gb}
  N.~R.~Constable and R.~C.~Myers,
  JHEP {\bf 9910} (1999) 037
  [hep-th/9908175].

\bibitem{Brower:2000rp}
  R.~C.~Brower, S.~D.~Mathur and C.~I.~Tan,
  Nucl.\ Phys.\ B {\bf 587} (2000) 249
  [hep-th/0003115].


\bibitem{Vento:2017ice}
  V.~Vento,
  Eur.\ Phys.\ J.\ A {\bf 53} (2017) no.9,  185
  doi:10.1140/epja/i2017-12378-2
  [arXiv:1706.06811 [hep-ph]].



\bibitem{Bali:1993fb}
  G.~S.~Bali {\it et al.} [UKQCD Collaboration],
  Phys.\ Lett.\ B {\bf 309} (1993) 378
  [hep-lat/9304012].

\bibitem{Morningstar:1999rf}
  C.~J.~Morningstar and M.~J.~Peardon,
  Phys.\ Rev.\ D {\bf 60} (1999) 034509
  [hep-lat/9901004].


\bibitem{Vaccarino:1999ku}
  A.~Vaccarino and D.~Weingarten,
  Phys.\ Rev.\ D {\bf 60} (1999) 114501
  [hep-lat/9910007].


\bibitem{Lee:1999kv}
  W.~J.~Lee and D.~Weingarten,
  Phys.\ Rev.\ D {\bf 61} (2000) 014015
  doi:10.1103/PhysRevD.61.014015
  [hep-lat/9910008].

\bibitem{Bali:2000vr}
  G.~S.~Bali {\it et al.} [TXL and T(X)L Collaborations],
  Phys.\ Rev.\ D {\bf 62} (2000) 054503
  doi:10.1103/PhysRevD.62.054503
  [hep-lat/0003012].

\bibitem{Hart:2001fp}
  A.~Hart {\it et al.} [UKQCD Collaboration],
  Phys.\ Rev.\ D {\bf 65} (2002) 034502
  doi:10.1103/PhysRevD.65.034502
  [hep-lat/0108022].


\bibitem{Lucini:2004my}
  B.~Lucini, M.~Teper and U.~Wenger,
  JHEP {\bf 0406} (2004) 012
  doi:10.1088/1126-6708/2004/06/012
  [hep-lat/0404008].


\bibitem{Chen:2005mg}
  Y.~Chen {\it et al.},
  Phys.\ Rev.\ D {\bf 73} (2006) 014516
  [hep-lat/0510074].


\bibitem{Gregory:2012hu}
  E.~Gregory, A.~Irving, B.~Lucini, C.~McNeile, A.~Rago, C.~Richards and E.~Rinaldi,
  JHEP {\bf 1210} (2012) 170
  [arXiv:1208.1858 [hep-lat]].
  
\bibitem{Lucini:2001ej}
  B.~Lucini and M.~Teper,
  JHEP {\bf 0106} (2001) 050
  doi:10.1088/1126-6708/2001/06/050
  [hep-lat/0103027].


\bibitem{Polchinski:2000uf}
  J.~Polchinski and M.~J.~Strassler,
  hep-th/0003136.


\bibitem{Brodsky:2003px}
  S.~J.~Brodsky and G.~F.~de Teramond,
  Phys.\ Lett.\ B {\bf 582} (2004) 211
  [hep-th/0310227].
  
\bibitem{Erlich:2005qh}
  J.~Erlich, E.~Katz, D.~T.~Son and M.~A.~Stephanov,
  Phys.\ Rev.\ Lett.\  {\bf 95} (2005) 261602
  [hep-ph/0501128].

\bibitem{DaRold:2005mxj}
  L.~Da Rold and A.~Pomarol,
  Nucl.\ Phys.\ B {\bf 721} (2005) 79
  [hep-ph/0501218].


\bibitem{Karch:2006pv}
  A.~Karch, E.~Katz, D.~T.~Son and M.~A.~Stephanov,
  Phys.\ Rev.\ D {\bf 74} (2006) 015005
  [hep-ph/0602229].
  
  
  \bibitem{reggenew}
  E.~Folco Capossoli and H.~Boschi-Filho,
  Phys.\ Lett.\ B {\bf 753}, 419 (2016)
  
  
  
  

\bibitem{BoschiFilho:2002vd}
  H.~Boschi-Filho and N.~R.~F.~Braga,
  JHEP {\bf 0305} (2003) 009
  doi:10.1088/1126-6708/2003/05/009
  [hep-th/0212207].

\bibitem{BoschiFilho:2005yh}
  H.~Boschi-Filho, N.~R.~F.~Braga and H.~L.~Carrion,
  Phys.\ Rev.\ D {\bf 73} (2006) 047901
  [hep-th/0507063].


\bibitem{Colangelo:2007pt}
  P.~Colangelo, F.~De Fazio, F.~Jugeau and S.~Nicotri,
  Phys.\ Lett.\ B {\bf 652} (2007) 73
  [hep-ph/0703316].


\bibitem{Forkel:2007ru}
  H.~Forkel,
  Phys.\ Rev.\ D {\bf 78} (2008) 025001
  [arXiv:0711.1179 [hep-ph]].
  

\bibitem{Li:2013pta}
  X.~F.~Li and A.~Zhang,
  Chin.\ Phys.\ C {\bf 38} (2014) no.1,  013102
  [arXiv:1309.7154 [hep-ph]].

\bibitem{Li:2013oda}
  D.~Li and M.~Huang,
  JHEP {\bf 1311} (2013) 088
  [arXiv:1303.6929 [hep-ph]].
  
\bibitem{Andreev:2006ct}
  O.~Andreev and V.~I.~Zakharov,
  Phys.\ Rev.\ D {\bf 74} (2006) 025023
  doi:10.1103/PhysRevD.74.025023
  [hep-ph/0604204].
  
 

\bibitem{White:2007tu}
  C.~D.~White,
  Phys.\ Lett.\ B {\bf 652} (2007) 79
  doi:10.1016/j.physletb.2007.07.006
  [hep-ph/0701157].

\bibitem{Bruni:2018dqm}
  R.~C.~L.~Bruni, E.~Folco Capossoli and H.~Boschi-Filho,
  arXiv:1806.05720 [hep-th].

\bibitem{Brodsky:2007hb}
  S.~J.~Brodsky and G.~F.~de Teramond,
  Phys.\ Rev.\ D {\bf 77} (2008) 056007
  doi:10.1103/PhysRevD.77.056007
  [arXiv:0707.3859 [hep-ph]].


\bibitem{Vento:2004xx}
  V.~Vento,
  Phys.\ Rev.\ D {\bf 73} (2006) 054006
  doi:10.1103/PhysRevD.73.054006
  [hep-ph/0401218].

  
\end{thebibliography}
\end{document}